\documentclass[twoside,twocolumn,9pt]{article}
\usepackage{extsizes}
\usepackage[super,sort&compress,comma,numbers]{natbib} 
\usepackage[version=3]{mhchem}
\usepackage[left=1.5cm, right=1.5cm, top=1.785cm, bottom=2.0cm]{geometry}
\usepackage{balance}
\usepackage{times, mathptmx}
\usepackage{sectsty}
\usepackage{graphicx}
\usepackage{lastpage}
\usepackage[format=plain,justification=justified,singlelinecheck=false,font={stretch=1.125,small,sf},labelfont=bf,labelsep=space]{caption}
\usepackage{float}
\usepackage{fancyhdr}
\usepackage{fnpos}
\usepackage[english]{babel}
\addto{\captionsenglish}{%
  
}
\usepackage{array}
\usepackage{droidsans}
\usepackage{charter}
\usepackage[T1]{fontenc}
\usepackage[usenames,dvipsnames,table]{xcolor}
\usepackage{setspace}
\usepackage[compact]{titlesec}
\usepackage{hyperref}
\usepackage{tabularx}
\usepackage{dblfloatfix}
\usepackage{transparent}
\usepackage[usetitle]{rsc}
\usepackage{epstopdf}%This line makes .eps figures into .pdf - please comment out if not required.
\usepackage{blindtext}

\definecolor{cream}{RGB}{222,217,201}

\begin{document}

\pagestyle{fancy}
\thispagestyle{plain}
\fancypagestyle{plain}{
%%%HEADER%%%
\renewcommand{\headrulewidth}{0pt}
}
%%%END OF HEADER%%%

%%%PAGE SETUP - Please do not change any commands within this section%%%
\makeFNbottom
\makeatletter
\renewcommand\LARGE{\@setfontsize\LARGE{15pt}{17}}
\renewcommand\Large{\@setfontsize\Large{12pt}{14}}
\renewcommand\large{\@setfontsize\large{10pt}{12}}
\renewcommand\footnotesize{\@setfontsize\footnotesize{7pt}{10}}
\makeatother

\renewcommand{\thefootnote}{\fnsymbol{footnote}}
\renewcommand\footnoterule{\vspace*{1pt}% 
\color{cream}\hrule width 3.5in height 0.4pt \color{black}\vspace*{5pt}} 
\setcounter{secnumdepth}{5}

\makeatletter 
\renewcommand\@biblabel[1]{#1}            
\renewcommand\@makefntext[1]% 
{\noindent\makebox[0pt][r]{\@thefnmark\,}#1}
\makeatother 
\renewcommand{\figurename}{\small{Fig.}~}
\sectionfont{\sffamily\Large}
\subsectionfont{\normalsize}
\subsubsectionfont{\bf}
\setstretch{1.125} %In particular, please do not alter this line.
\setlength{\skip\footins}{0.8cm}
\setlength{\footnotesep}{0.25cm}
\setlength{\jot}{10pt}
\titlespacing*{\section}{0pt}{4pt}{4pt}
\titlespacing*{\subsection}{0pt}{15pt}{1pt}
%%%END OF PAGE SETUP%%%

%%%FOOTER%%%
\fancyfoot{}
\fancyfoot[LO,RE]{\vspace{-7.1pt}\includegraphics[height=9pt]{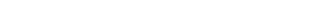}}
\fancyfoot[CO]{\vspace{-7.1pt}\hspace{13.2cm}\includegraphics{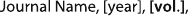}}
\fancyfoot[CE]{\vspace{-7.2pt}\hspace{-14.2cm}\includegraphics{head_foot/RF.pdf}}
\fancyfoot[RO]{\footnotesize{\sffamily{1--\pageref{LastPage} ~\textbar  \hspace{2pt}\thepage}}}
\fancyfoot[LE]{\footnotesize{\sffamily{\thepage~\textbar\hspace{3.45cm} 1--\pageref{LastPage}}}}
\fancyhead{}
\renewcommand{\headrulewidth}{0pt} 
\renewcommand{\footrulewidth}{0pt}
\setlength{\arrayrulewidth}{1pt}
\setlength{\columnsep}{6.5mm}
\setlength\bibsep{1pt}
%%%END OF FOOTER%%%

%%%FIGURE SETUP - please do not change any commands within this section%%%
\makeatletter 
\newlength{\figrulesep} 
\setlength{\figrulesep}{0.5\textfloatsep} 

\newcommand{\topfigrule}{\vspace*{-1pt}% 
\noindent{\color{cream}\rule[-\figrulesep]{\columnwidth}{1.5pt}} }

\newcommand{\botfigrule}{\vspace*{-2pt}% 
\noindent{\color{cream}\rule[\figrulesep]{\columnwidth}{1.5pt}} }

\newcommand{\dblfigrule}{\vspace*{-1pt}% 
\noindent{\color{cream}\rule[-\figrulesep]{\textwidth}{1.5pt}} }

\makeatother
%%%END OF FIGURE SETUP%%%

%%%TITLE, AUTHORS AND ABSTRACT%%%
\twocolumn[
  \begin{@twocolumnfalse}
{\includegraphics[height=30pt]{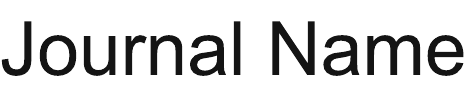}\hfill\raisebox{0pt}[0pt][0pt]{\includegraphics[height=55pt]{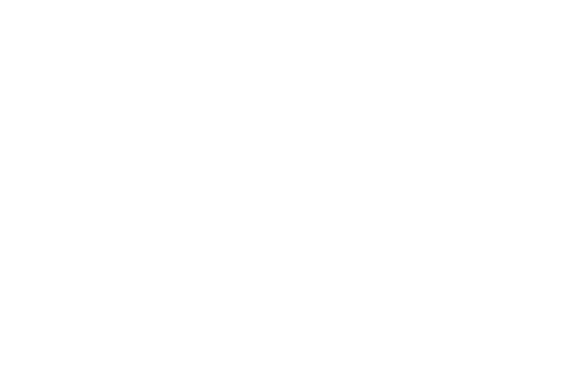}}\\[1ex]
\includegraphics[width=18.5cm]{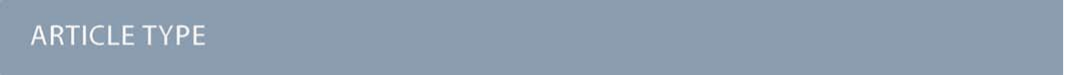}}\par
\vspace{1em}
\sffamily
\begin{tabular}{m{4.5cm} p{13.5cm} }

\includegraphics{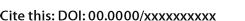} & \noindent\LARGE{\textbf{Increasing the Collection Efficiency in Selenium \newline Thin-Film Solar Cells Using a Closed-Space Annealing Strategy}} \\%Article title goes here instead of the text "This is the title"
\vspace{0.3cm} & \vspace{0.3cm} \\

 & \noindent\large{Rasmus Nielsen,$^{\ast}$\textit{$^{a}$} Markus Schleuning,\textit{$^{b}$} Orestis Karalis,\textit{$^{b}$} Tobias H. Hemmingsen,\textit{$^{a}$} Ole Hansen,\textit{$^{c}$} Ib Chorkendorff,\textit{$^{a}$} Thomas Unold,\textit{$^{b}$} and Peter C. K. Vesborg\textit{$^{a}$}} \\%Author names go here instead of "Full name", etc.

\includegraphics{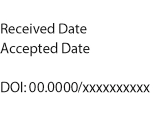} & \noindent\normalsize{Elemental selenium is a promising wide-bandgap ($E_\mathrm{G}\approx\text{1.95}$ eV) photovoltaic material for the next generation of thin-film solar cells. To realize high-efficiency selenium solar cells, it is crucial to optimize the crystallization process of the selenium thin-film photoabsorber. However, the high vapor pressure of selenium restricts the processing conditions to a compromise between the growth of large crystal grains and the formation of pinholes. In this study, we introduce a closed-space annealing (CSA) strategy designed to suppress the sublimation of selenium, enabling thermal annealing processes at higher temperatures and for longer periods of time. As a result, we consistently improve carrier collection and the overall photovoltaic device performance in our selenium solar cells. By characterizing the carrier dynamics in our devices, we conclude that the observed improvements result from a reduction in charge transfer resistance rather than an increase in carrier diffusion length. The CSA strategy is a promising method for controlling surface morphology and roughness without reducing crystal grain sizes, which paves the way for further advancements in the efficiency and reproducibility of selenium thin-film solar cells.} \\%The abstrast goes here instead of the text "The abstract should be..."
\end{tabular}

 \end{@twocolumnfalse} \vspace{0.6cm}

  ]
%%%END OF TITLE, AUTHORS AND ABSTRACT%%%

%%%FONT SETUP - please do not change any commands within this section
\renewcommand*\rmdefault{bch}\normalfont\upshape
\rmfamily
\section*{}
\vspace{-1cm}

%%%FOOTNOTES%%%

\footnotetext{\textit{$^{a}$~SurfCat, DTU Physics, Technical University of Denmark, DK-2800 Kgs. Lyngby, Denmark.}}
\footnotetext{\textit{$^{b}$~Department of Structure and Dynamics of Energy Materials, Helmholtz-Zentrum Berlin f\"ur Materialien und Energie GmbH, Berlin, Germany.}}
\footnotetext{\textit{$^{c}$~DTU Nanolab, National Center for Nano Fabrication and Characterization, Technical University of Denmark, DK-2800, Kgs. Lyngby, Denmark.}}
\footnotetext{\textit{$^{\ast}$~ E-mail: raniel@dtu.dk}}

%Please use \dag to cite the ESI in the main text of the article.
%If you article does not have ESI please remove the the \dag symbol from the title and the footnotetext below.
\footnotetext{\dag~Electronic Supplementary Information (ESI) available: Additional figures. See DOI: 00.0000/xxxxxxxxxx/}
%additional addresses can be cited as above using the lower-case letters, c, d, e... If all authors are from the same address, no letter is required

%\footnotetext{\ddag~Additional footnotes to the title and authors can be included \textit{e.g.}\ `Present address:' or `These authors contributed equally to this work' as above using the symbols: \ddag, \textsection, and \P. Please place the appropriate symbol next to the author's name and include a \texttt{\textbackslash footnotetext} entry in the the correct place in the list.}

%%%END OF FOOTNOTES%%%

%%%MAIN TEXT%%%%

\section*{INTRODUCTION}
Selenium is experiencing renewed interest as an inorganic photovoltaic material for the next generation of thin-film solar cells\cite{mitzi2022a, zhu2019a}. With a direct bandgap of 1.95 eV and high absorption coefficient in its trigonal phase\cite{tutihasi1967a}, selenium is potentially an ideal candidate for the photoabsorbing layer in indoor solar cells\cite{yan2022a, ieee2017a, wei2023a}, as well as the top cell in tandem devices\cite{youngman2021a, nielsen2023a}. Furthermore, as selenium crystallizes at approximately 120$^\circ$C and melts at 220$^\circ$C\cite{lu2022a}, the temperatures needed for processing are relatively low in comparison with other chalcogenide-based absorber materials\cite{hadke2022a}. This makes selenium compatible with most substrates and bottom cell candidates, and facilitates the production of low-cost solar cells.

Selenium thin-films are predominantly synthesized by evaporating an amorphous layer of selenium, followed by a thermal annealing process to crystallize the as-deposited film\cite{nielsen2021a, fu2022a, zheng2022a, nakada1985a, wang2014a}. However, the temperature window leading to the highest power conversion efficiencies is fairly narrow\cite{hadar2019a}, and the high vapor pressure of selenium poses challenges in growing high-quality crystalline selenium thin-films without compromising the surface morphology and forming pinholes\cite{nielsen2023b, todorov2017a}.

In 2017, D. M. Bishop et al. demonstrated that annealing at suboptimal temperatures significantly reduces the collection efficiency of charge carriers excited by longer wavelength photons\cite{ieee2017a}. They attributed this result to a reduction in the carrier diffusion length, as these charge carriers are generated furthest from the carrier-separating pn-junction. In 2019, I. Hadar et al. supplemented this study by showing that the surface morphology changes drastically as a function of the annealing temperature, particularly within the range of 190$^\circ$C to 195$^\circ$C\cite{hadar2019a}. Building on these findings, we reported a record fill-factor FF=63.7\% for selenium solar cells obtained by laser-annealing the absorber through the substrate, which resulted in a crystalline selenium thin-film with a negligible surface roughness\cite{nielsen2023b}. These results indicate that the collection efficiency is strongly influenced by the annealing temperature and possibly the surface morphology.

\begin{figure*}[t!]
    \centering
    \includegraphics[width=\textwidth]{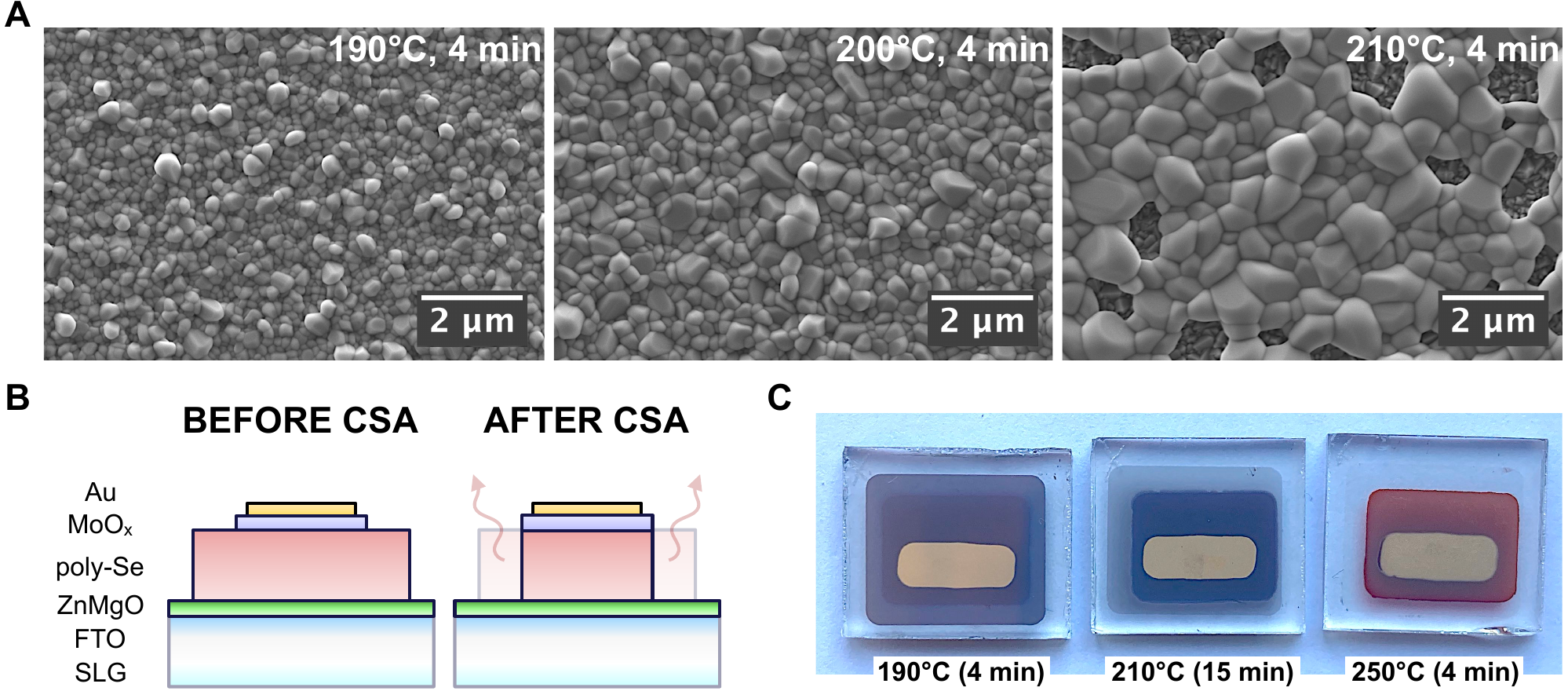}
    \caption{(\textbf{A}) Top-view SEM images of selenium thin-films deposited on ZnMgO and crystallized for 4 minutes at 190$^\circ$C, 200$^\circ$C, and 210$^\circ$C, respectively. (\textbf{B}) Schematic representation of a selenium solar cell before and after CSA, illustrating the sublimation of unprotected edges of the selenium thin-film from the substrate. (\textbf{C}) Optical photograph depicting three selenium solar cells after CSA post-processing using different annealing conditions.}
    \label{fig:Figure1}
\end{figure*}

In this work, we introduce a closed-space annealing (CSA) strategy wherein the selenium thin-film is encapsulated with an ultra-thin layer of MoO$_\text{x}$. This layered structure can be thermally annealed at temperatures well above the melting point of selenium without compromising the integrity of the photoabsorber. As MoO$_\text{x}$ has also been established as an excellent hole-selective contact material in selenium solar cells, this encapsulant holds its relevance in the context of photovoltaic devices. Consequently, we fabricate selenium solar cells and demonstrate that the CSA-strategy significantly enhances carrier collection, thereby improving the overall photovoltaic device performance. Using time-resolved terahertz spectroscopy (TRTS), transient surface photovoltage (t-SPV), and impedance spectroscopy, we show that this notable improvement is not a result of increased carrier diffusion lengths but rather a reduction in charge transfer resistance.\\

\section*{RESULTS AND DISCUSSION}

Previous studies have highlighted the sensitivity of the crystallization process of selenium to the substrate on which the selenium thin-film is deposited\cite{hadar2019a, todorov2017a}. Therefore, the annealing conditions should be carefully optimized for different device architectures. As the selenium solar cells presented in this study comprise the structure SLG/FTO/ZnMgO/poly-Se/MoO$_\text{x}$/Au, we investigated the effect of different annealing temperatures on the crystallization of amorphous selenium deposited on ZnMgO. Figure \ref{fig:Figure1}A shows top-view scanning electron microscopy (SEM) images of selenium thin-films crystallized by thermal annealing in a preheated aluminium oven for 4 minutes at temperatures of 190$^\circ$C, 200$^\circ$C, and 210$^\circ$C, respectively. We observed significantly larger crystal grains with increasing annealing temperatures; however, at 210$^\circ$C, large pinholes are visible in the absorber thin-film, which we believe is a result of the sublimation of selenium. Therefore, to fabricate high-quality selenium absorbers with large crystal grains, the sublimation of selenium at higher annealing temperatures must be suppressed.

We hypothesized that by forming a closed space through which selenium cannot escape, annealing processes may be carried out at higher temperatures or for longer periods of time. Similar closed-space annealing strategies have been demonstrated to improve the optoelectronic quality of different perovskite materials and reduce the density of grain boundaries as smaller adjacent crystals merge into larger ones\cite{wang2022a}. For the selenium thin-film to serve as the absorber in a photovoltaic device, the encapsulating material must either be sacrificial or serve as a carrier-selective contact layer. As MoO$_\text{x}$ is used as a hole-transport layer in the best performing selenium solar cells reported to date\cite{todorov2017a, nielsen2022a}, we investigated its candidacy as an encapsulant.

To assess the effectiveness of MoO$_\text{x}$ as an encapsulating material, we deposited each thin-film in our devices with successively smaller aperture areas, as illustrated in Figure \ref{fig:Figure1}B. Our previous work highlighted the necessity of crystallizing the as-deposited selenium thin-film prior to the deposition of MoO$_\text{x}$ to avoid the formation of cracks in the absorber\cite{nielsen2023a}. Therefore, we annealed the SLG/FTO/ZnMgO/a-Se structure for 4 minutes at 190$^\circ$C to transform the amorphous selenium (a-Se) into a poly-crystalline selenium (poly-Se) thin-film of the desired trigonal phase, followed by the deposition of MoO$_\text{x}$ and the gold electrode. Given that the MoO$_\text{x}$ thin-film only partially covers the poly-Se thin-film, we expect the unprotected edges of the absorber to sublimate during higher temperature annealing processes. This expectation is in agreement with the visual observations made from the photograph in Figure \ref{fig:Figure1}C, showing three devices after CSA post-processing using different annealing conditions. 

\begin{figure*}[t!]
    \centering
    \includegraphics[width=0.66\textwidth]{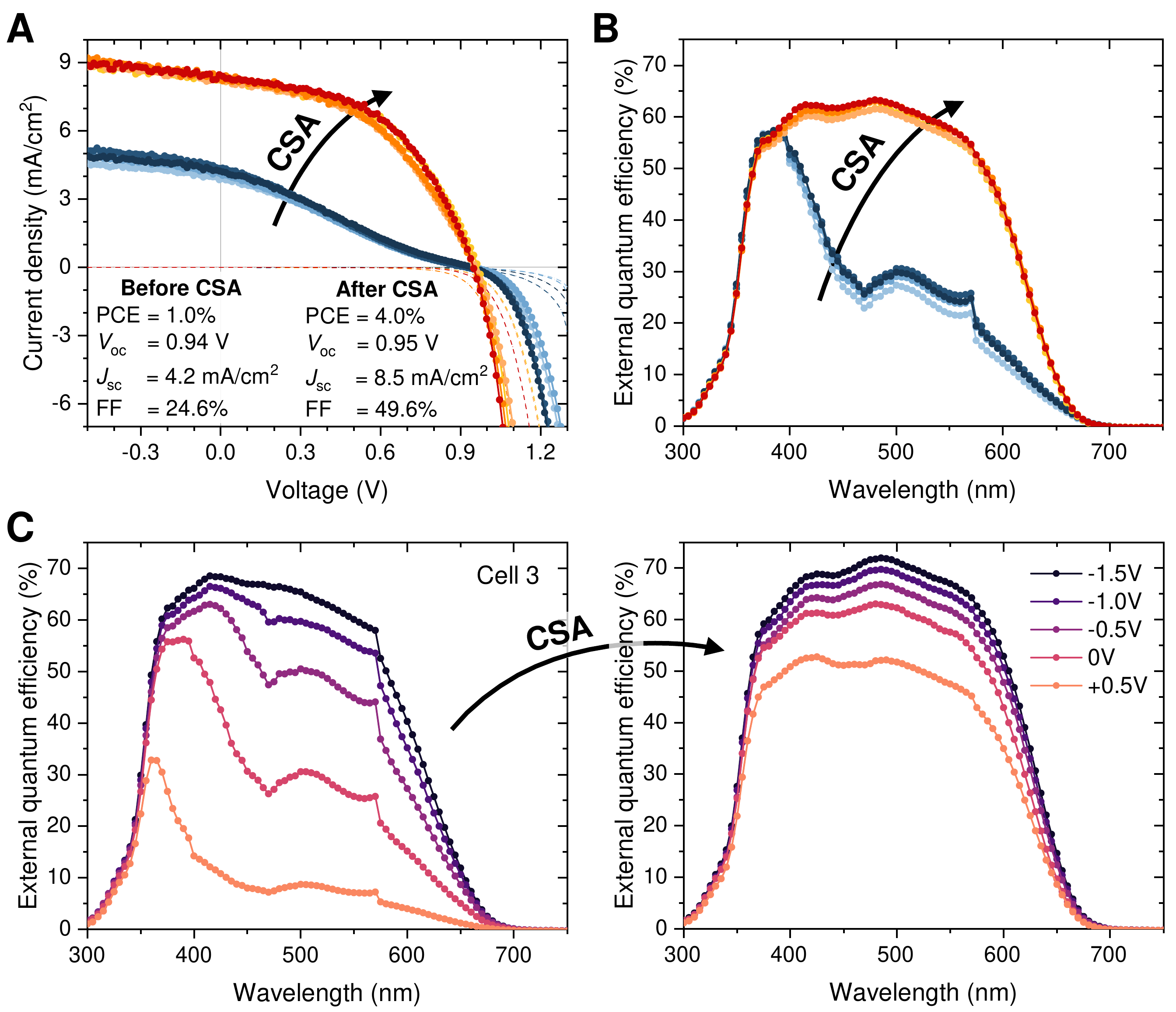}
    \caption{Photovoltaic device performance of a batch of selenium thin-film solar cells before and after a CSA post-processing step carried out at 210$^\circ$C for 15 minutes. (\textbf{A}) Current-voltage (JV) curves measured under dark (dashed lines) and 1 Sun illuminated (solid lines) conditions. (\textbf{B}) External quantum efficiency (EQE) spectra measured under short-circuit conditions. (\textbf{C}) EQE-spectra measured while applying an external bias to the device.}
    \label{fig:Figure2}
\end{figure*}

Figure \ref{fig:Figure2}A and B show the current-voltage (JV) curves and external quantum efficiency (EQE) spectra of a batch of devices before and after a CSA post-processing step at 210$^\circ$C for 15 minutes. Before CSA, we observe a pronounced S-shaped kink in the JV-curves and reduced collection efficiencies at longer wavelengths in the EQE-spectra. These EQE-spectra qualitatively resemble the findings published by D. M. Bishop et al.\cite{ieee2017a}, suggesting that a crystallization temperature of 190$^\circ$C is suboptimal. However, after CSA, the carrier collection at lower photon energies improves substantially, and the JV-curves no longer exhibit the S-shaped kink. Consequently, the CSA post-processing step doubled the short-circuit current densities and increased the power conversion efficiencies (PCE) by a factor of $\sim$4.

S-shaped kinks in JV-curves are often encountered in the development of new solar cells, and are typically reported to be the result of charge transport barriers at the carrier-selective contacts\cite{saive2019a}. To gain a deeper understanding of this voltage-dependent non-ideality, the EQE-spectra in Figure \ref{fig:Figure2}C were measured while applying an external bias to the device. Before CSA, the reductions in carrier collection with increasing externally applied bias are observed to be wavelength-dependent. In general, a wavelength-dependent reduction in quantum efficiency is the result of poor collection of minority carriers generated deeper in the absorber, as these carriers have to diffuse further before reaching the depletion region. In this case, a reverse bias will improve carrier collection as the width of the depletion region increases, and as the penetration depth is deeper for lower energy photons, the effect is more dramatic for carriers generated at longer wavelengths. However, after CSA, the reduction in quantum efficiency is observed to be wavelength-independent. In this case, the carrier collection is limited by mechanisms that affect all carriers equally regardless of where in the device they are generated. Such mechanisms include interface recombination, recombination in the depletion region, transport barriers, or recombination dominated by band tailing\cite{hages2016a}. Therefore, it would appear as if the carrier diffusion length has improved.

\begin{figure*}
    \centering
    \includegraphics[width=\textwidth]{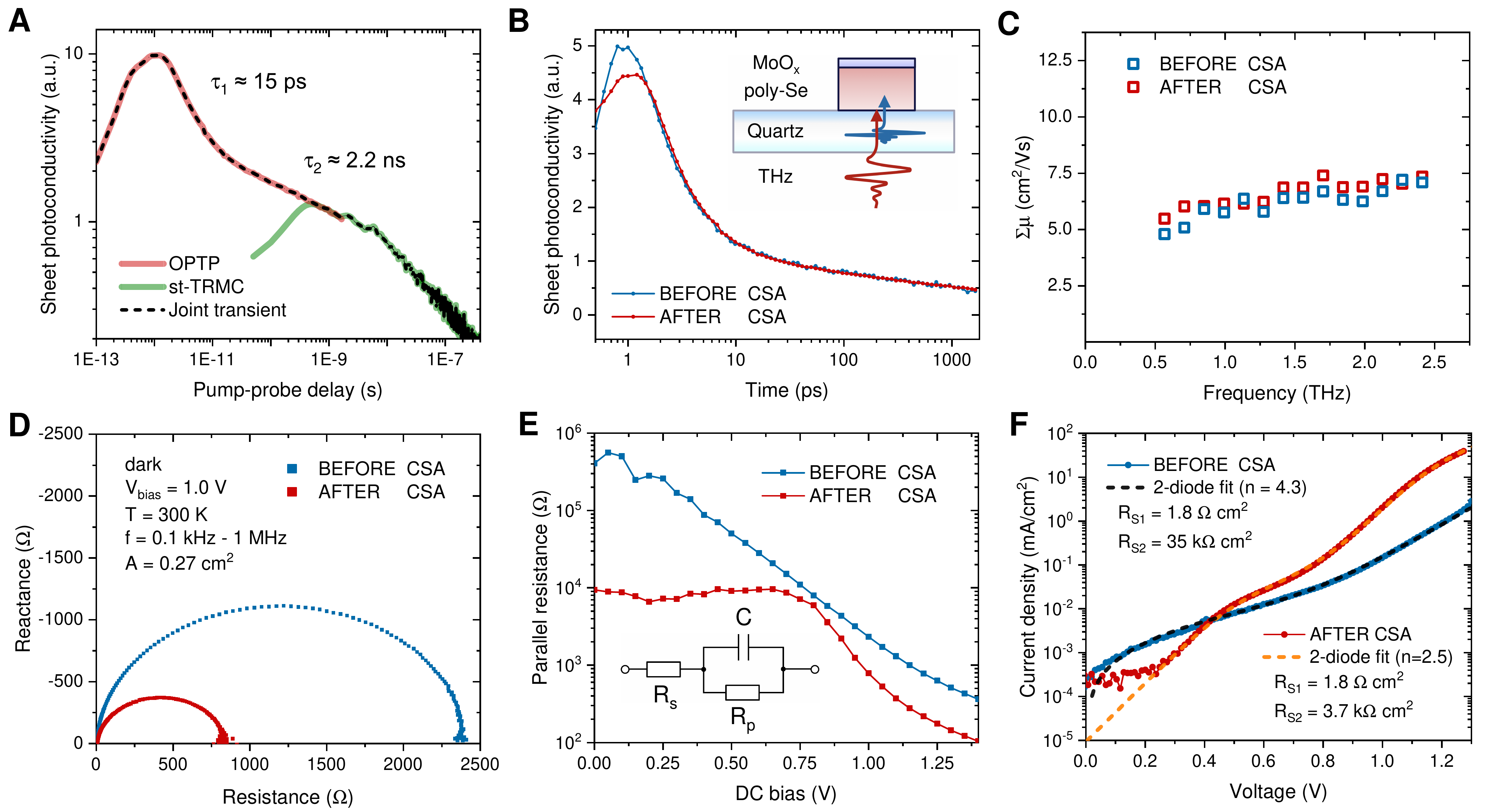}
    \caption{(\textbf{A}) Joint sheet photoconductivity transients measured using optical-pump terahertz-probe (OPTP) spectroscopy and sample-terminated time-resolved microwave conductivity (st-TRMC). (\textbf{B}) OPTP-transients measured on the layered poly-Se/MoO$_\text{x}$ structure synthesized on quartz. (\textbf{C}) The corresponding AC-mobility spectrum of poly-Se. (\textbf{D}) Nyquist plot measured under dark conditions at a bias voltage of 1.0 V. (\textbf{E}) The parallel resistance fitted to dark impedance spectra measured various DC biases using a Randles circuit. (\textbf{F}) Dark JV-curves fitted to a two-diode model.}
    \label{fig:Figure3}
\end{figure*}

%To determine the carrier diffusion length in the selenium thin-film before and after CSA, we obtained carrier mobilities and lifetimes from photoconductivity transients measured using optical-pump terahertz-probe (OPTP) spectroscopy and sample-terminated time-resolved microwave conductivity (st-TRMC). These measurements were carried out on the layered poly-Se/MoO$_\text{x}$ structure synthesized directly on quartz substrates. The joint sheet photoconductivity transient shown in Figure \ref{fig:Figure3}A is fitted to a biexponential function to retrieve the two characteristic lifetimes, $\tau_\text{1}=\text{15}$ ps and $\tau_\text{2}=\text{2.2}$ ns, respectively. Considering the open-circuit voltage deficit is $\approx$700 mV, we do not expect an effective carrier lifetime on the order of several nanoseconds\cite{kirchartz2018a, kirchartz2018b}. Therefore, we interpret the rapid initial decay as the carrier lifetime. However, as shown in Figure \ref{fig:Figure3}B, the transient decay remains unchanged before and after CSA. Figure \ref{fig:Figure3}C shows the AC-mobility sum derived from the pump-induced changes in the transmitted THz probe, extrapolated to a DC-value of $\Sigma \mu\approx\text{5}$ cm$^\text{2}$/Vs both before and after CSA. Surprisingly, this suggests that the mobility-lifetime product, and thus the carrier diffusion length, has not been improved by CSA.

To determine the carrier diffusion length in the selenium thin-film before and after CSA, we obtained carrier mobilities and lifetimes from photoconductivity transients measured using optical-pump terahertz-probe (OPTP) spectroscopy and sample-terminated time-resolved microwave conductivity (st-TRMC). These measurements were carried out on the layered poly-Se/MoO$_\text{x}$ structure synthesized directly on quartz substrates. The joint sheet photoconductivity transient shown in Figure \ref{fig:Figure3}A is fitted to a biexponential function to retrieve the two characteristic lifetimes, $\tau_\text{1}=\text{15}$ ps and $\tau_\text{2}=\text{2.2}$ ns, respectively. Considering the open-circuit voltage deficit of $\approx$700 mV, we do not expect an effective carrier lifetime on the order of several nanoseconds\cite{kirchartz2018a, kirchartz2018b}, but a carrier lifetime on the order of picosecond would imply unreasonably short carrier diffusion lengths of approximately ten nanometers. Regardless of how the carrier lifetime is interpreted from the photoconductance decay, the transient remains unchanged before and after CSA, as shown in Figure \ref{fig:Figure3}B. Figure \ref{fig:Figure3}C shows the AC-mobility sum derived from the pump-induced changes in the transmitted THz probe, extrapolated to a DC-value of $\Sigma \mu\approx\text{5}$ cm$^\text{2}$/Vs both before and after CSA. Surprisingly, this suggests that the mobility-lifetime product, and thus the carrier diffusion length, has not been improved by CSA.

To further characterize the influence of CSA on the photovoltaic devices, we carried out impedance spectroscopy measurements in the dark with various applied DC biases. The Nyquist plots in Figure \ref{fig:Figure3}D show a smaller semicircle after CSA. By modeling the selenium solar cell using a Randles-circuit, the parallel resistor quantifies the diameter of these semicircles. We observe that the parallel resistance in Figure \ref{fig:Figure3}E, typically attributed to charge transfer resistance and recombination losses, is smaller at all applied DC biases after CSA. However, these fitted values appear to be voltage-independent at lower applied DC biases. This plateau may be explained by fitting the dark JV-curves in Figure \ref{fig:Figure3}F using two parallel-connected diodes, where the series resistance of the secondary diode, $\text{R}_\text{S2}$, is on the order of $\sim10^5 \Omega$ before, and $\sim10^4 \Omega$ after CSA. This implies that as the frequency increases during impedance spectroscopy, the capacitor of the secondary diode represents an increasingly smaller impedance, eventually equivalent to a short-circuit. Therefore, the AC-response of the device may be shorted by a $\sim10^4 \Omega$ resistor after CSA, matching the order of magnitude observed for the voltage-independent region of the fitted parallel resistance. Another important point to note is the substantial improvement in the diode ideality factor in the dark, indicating an overall healthier device.

\begin{figure*}
    \centering
    \includegraphics[width=\textwidth]{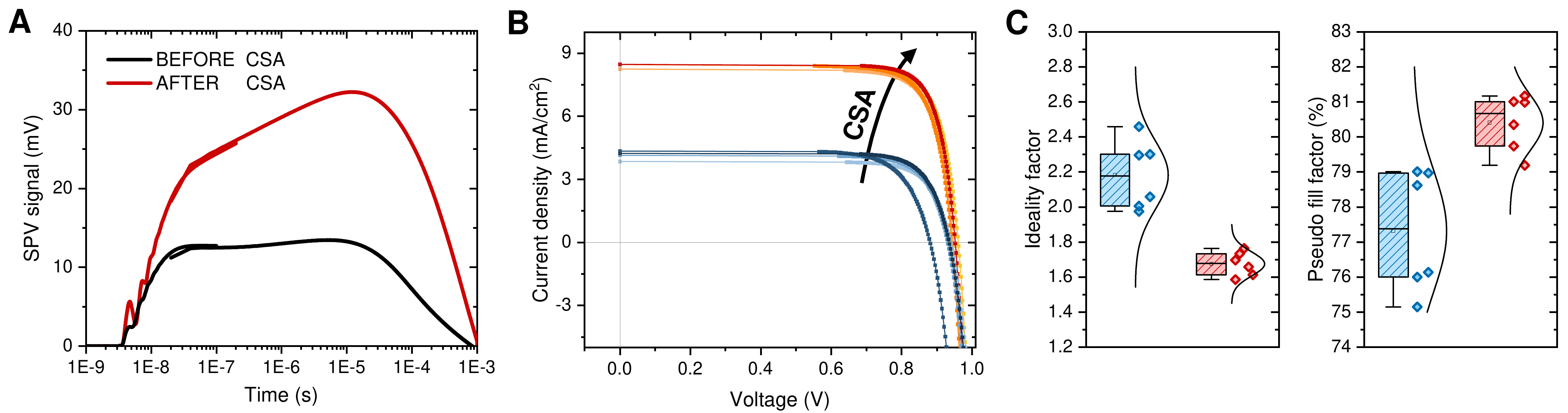}
    \caption{(\textbf{A}) Surface photovoltage (SPV) transients of complete selenium solar cells excited using a 515 nm laser, at a repetition rate of $f_\text{rep}$=1 kHz. (\textbf{B}) Pseudo JV-curves reconstructed from Suns-V$_\text{oc}$ measurements. (\textbf{C}) Batch statistics showing the improved ideality factor and pseudo fill factor.}
    \label{fig:Figure4}
\end{figure*}

As the carrier lifetimes and open-circuit voltages are similar for all devices before and after CSA, it is more likely that CSA is improving charge transfer resistance rather than suppressing the dominant carrier recombination processes. Therefore, we investigated carrier extraction using transient surface photovoltage (t-SPV) on our photovoltaic devices. To identify the different dynamic processes, the t-SPV signal is recorded over a wide time range. As shown in Figure \ref{fig:Figure4}A, the overall magnitude of the transient signal increases after CSA, indicating a more efficient transfer of charge carriers to the respective contacts, which is in good agreement with our EQE-spectra. The initial rapid rise within the first 20 ns of the optical pump is typically assigned to the transfer of carriers from the absorber to the contacts; in this case, Au and FTO. After CSA, the SPV signal continues to rise. The dynamic processes usually associated with a timescale of tens of microseconds is the de-trapping of charge carriers\cite{levine2021a, neukom2018a, neukom2019a}, but the complex carrier kinetics caused by emission and recombination processes via shallow defects is quite intricate. Therefore, we content ourselves with the qualitative observation of an additional increase in carrier collection after CSA, possibly from the re-emission of carriers trapped in shallow defects. Finally, the slow relaxation of the SPV signal is assigned to the transfer of charge carriers from the contacts back into the absorber where they recombine, as well as the recombination of slowly de-trapped carriers.

To gain further insights into the health of the device under illuminated conditions, we carried out injection-dependent open-circuit voltage measurements, also known as Suns-V$_\text{oc}$. Figure \ref{fig:Figure4}B shows pseudo JV-curves recreated from Suns-V$_\text{oc}$ measurements. As no charge carriers are transported in the device during these measurements, the effects of series resistance are eliminated, allowing for a more reliable fit of the non-ideal diode equation. We find that the ideality factor under 1 Sun illumination improves from an average value of $\sim$2.2 to $\sim$1.7, with a smaller spread, as shown in Figure \ref{fig:Figure4}C. This implies that recombination in the depletion region has been slightly suppressed, improving the pseudo fill factor obtained from Suns-V$_\text{oc}$ from $\sim$77\% to $\sim$81\%. The difference between the actual fill factor and the pseudo fill factor is accounted for by series resistance and bias-dependent collection of charge carriers.

\begin{figure}
    \centering
    \includegraphics[width=0.47\textwidth]{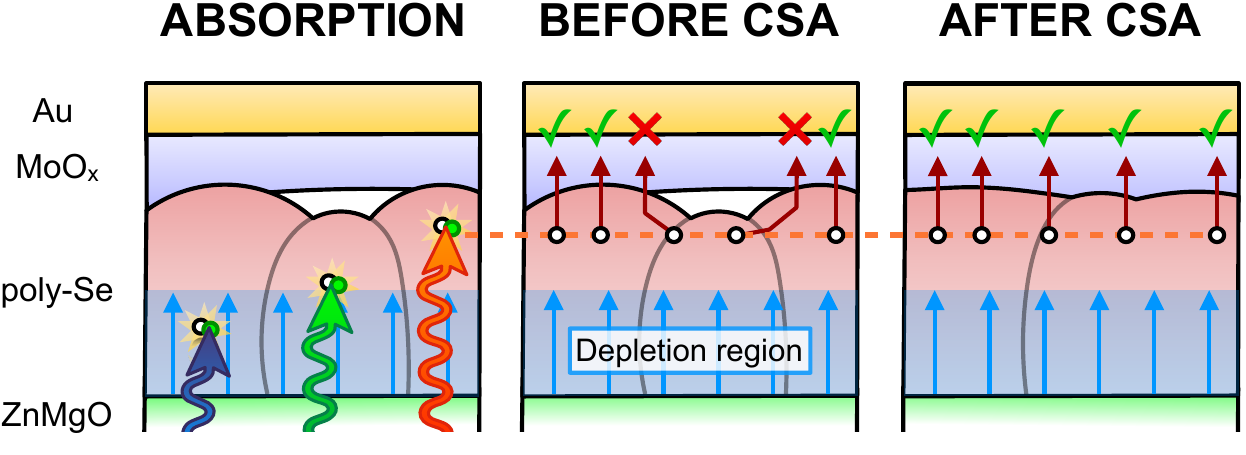}
    \caption{Schematic illustration of the wavelength-dependent photogeneration of charge carriers in the absorber, as well as partially delaminated contact layer before CSA, and how CSA is hypothesized to improve both the electrical contact and merge adjacent crystals into larger ones.}
    \label{fig:Figure5}
\end{figure}

In summary, we have observed an improved carrier collection from EQE-spectra under various applied biases, as well as through t-SPV measurements. The wavelength-dependent reductions in quantum efficiency with increasing bias are commonly attributed to poor carrier diffusion lengths. However, our findings from time-resolved terahertz spectroscopy measurements indicate that the diffusion length is not changed by the CSA treatment. Therefore, the observed reduction in parallel resistance, as revealed by impedance spectroscopy, is ascribed to a decrease in the charge transfer resistance. We hypothesize that the wavelength-dependence may be the result of poor electrical contact between one or more functional layers in the devices. As we approach the melting point of selenium, the absorber is expected to recrystallize, providing the carrier-selective contact films with a second chance to properly adhere to the absorber. This is schematically illustrated in Figure \ref{fig:Figure5}, showing how the lower energy photons penetrate deeper into the absorber. Assuming partially delaminated layers, carriers generated in the vicinity of a delamination must diffuse further to reach the carrier-selective contact. Therefore, by improving the electrical contact between the layers in the stack, more carriers may be collected without increasing the carrier diffusion length within the absorber material.

The CSA post-processing step introduced in this work proves to be an effective strategy for reducing charge transfer resistance losses in selenium thin-film solar cells. The observed improvements in carrier collection in the EQE and the elimination of S-shaped kink in the JV-curves shows that higher annealing temperatures lead to significantly higher power conversion efficiencies. As the CSA-strategy widens the temperature window in which the integrity of the selenium absorber layer is not compromised, it is a highly promising approach to improve the overall performance of selenium solar cells regardless of the device architecture. Further investigation is needed to determine optimal processing conditions for both the crystallization step and the final CSA step. Given that surface roughness decreases with lower crystallization temperatures and that CSA eliminates the performance limitations associated with suboptimal crystallization temperatures, CSA could potentially facilitate the synthesis of selenium solar cells with negligible roughness. However, thorough investigation of the combined parameter space of the crystallization step and the final CSA post-processing step is for future studies.\\

\section*{CONCLUSION}

In conclusion, we have introduced a closed-space annealing strategy (CSA) which effectively improves carrier collection and widens the temperature processing window of selenium solar cells. This allows for the selenium absorber layer to be crystallized at suboptimal temperatures, where the integrity of the thin-film is better preserved. During CSA, the absorber recrystallizes, unlocking efficient carrier collection and high power conversion efficiencies. We demonstrate promising photovoltaic performance improvements including an increase in the power conversion efficiency by a factor of 4, a doubling of the short-circuit current density, and an improved diode ideality factor under both dark and 1 Sun illuminated conditions.

Through transient photoconductance measurements, we observe that CSA does not influence the carrier lifetime or mobility. Consequently, we do not attribute the improved performance to an increase in carrier diffusion length. Instead, based on impedance spectroscopy and transient surface photovoltage, we find that the improved carrier collection possibly originates from a reduction in charge transfer resistance. One possible explanation is that the carrier-selective contact layers have delaminated, suggesting that charge carriers would have to diffuse laterally in the absorber to be collected. We propose that as the absorber recrystallizes during CSA, adjacent grains merging into larger ones, and the delaminated contacts are given a second chance to properly adhere to the absorber layer. The improved electrical contact between the layers in the stack effectively reduces charge transfer resistance, resulting in an increased carrier collection without improving carrier diffusion length.\\

\section*{METHODS}

\begin{small}

The selenium thin-film solar cells presented in this work comprise the device structure SLG/FTO/Zn$_\text{0.85}$Mg$_\text{0.15}$O/poly-Se/MoO$_\text{x}$/Au. The experimental details of the materials used and the fabrication process flow are available in a previous publication\cite{nielsen2022a}. 300 nm thick selenium thin-films encapsulated with 15 nm layer of MoO$_\text{x}$ are deposited on quartz substrates for THz and microwave conductivity measurements.\\

\paragraph*{Characterization of solar cells.~~} The current-voltage (JV) characteristics of the solar cells are measured in the dark and under 1 Sun illumination (1000 W/m$^\text{2}$, AM1.5G) using a Keithley 2561A source meter with a four-terminal sensing setup. The light source is an Oriel Sol2A Class ABA solar simulator from Newport equipped with a 1600W Xe arc lamp and optical filters, and the intensity is calibrated in the plane of the device under test using a reference solar cell from Orion. As no mask aperture is used during the acquisition, the short-circuit current density (J$_\text{sc}$) is calibrated using the AM1.5G-equivalent current density calculated by integrating the product of the external quantum efficiency (EQE) and the photon flux of the AM1.5G solar spectrum\cite{snaith2012a}. The EQE spectra are measured using the QEXL from PV Measurements calibrated using a silicon photodiode. Injection-level-dependent open-circuit voltage (Suns-V$_\text{oc}$) measurements are carried out using the WCT-120 from Sinton Instruments. The impedance spectra are measured using a BK895 LCR meter operating in resistance-reactance mode at room temperature in the dark. These measurements were performed in a frequency range from 0.1 kHz to 1 MHz with an AC level of 30 mV and DC biases ranging from -0.8 V to +1.4 V. The impedance spectra are fitted to a Randles-circuit using Elchemea Analytical. All scanning electron microscopy (SEM) images were obtained using a Supra 40 VP SEM from Zeiss.\\

\paragraph*{THz and microwave conductivity.~~} For the time-resolved terahertz spectroscopy (TRTS) and microwave conductivity (TRMC) measurements, the quartz/poly-Se/MoO$_\text{x}$ samples are optically pumped using laser pulses with a wavelength of 400 nm, a pulse width (FWHM) of $\sim$70 fs, and a repetition rate of 150 kHz. The photoexcited charge carriers absorb THz pulses generated by the optical rectification of 800 nm femtosecond laser pulses in a (110)-oriented ZnTe crystal. The pump-induced changes in the transmitted THz pulse are measured and analyzed for the sheet photoconductivity using the thin-film approximation\cite{hempel2022a}. The transient photoconductivity is obtained by sampling the peak amplitude of the transmitted THz probe and scanning the delay between the optical pump and THz pulse using a mechanical delay stage. The frequency-dependent carrier mobility sum $\Sigma\mu$ is obtained by sampling the entire THz pulse at a fixed pump-probe delay of 500 ps, achieved by shifting the transmitted signal using a second delay line. For the TRMC measurements, microwaves in the K$_\text{a}$ band (26-40 GHz) are generated by a gun diode with an adjustable internal resonator length and directed towards the sample surface using waveguides. Here, the signal is partially reflected, and the pump-induced changes to the phase and amplitude, which relate to the change in the sheet photoconductance, are detected using a Rohde and Schwarz RTO2044 oscilloscope. Additional details are available in a previous publication\cite{schleuning2022a}. \\

\paragraph*{Surface photovoltage.~~} The transient surface photovoltage (t-SPV) measurements are carried out on selenium solar cells under ambient conditions using laser pulses with a wavelength of 515 nm, a pulse width (FWHM) of $\sim$160 ps, a repetition rate of $f_\text{rep}$=1 kHz, and an optical energy density of 24.4 nJ/cm$^\text{2}$/pulse measured using a power meter. The photoexcitation is performed from the SLG/FTO side of the devices, and the laser fluence is controlled using neutral density filters. The SPV signal is acquired using a parallel plate capacitor configuration, comprising the Au-contact of the solar cells, a $\sim$0.5 mm sheet of mica as an insulator, and a quartz cylinder partially coated with an FTO electrode. Additional details are available in a previous publication\cite{levine2021a}.\\

\end{small}

\section*{Conflicts of interest}
There are no conflicts to declare.\\

\section*{Acknowledgements}
This work was supported by the Independent Research Fund Denmark (DFF) grant 0217-00333B, and the Villum Foundation V-SUSTAIN grant 9455.\\

\section*{Data availability}
The data that support the findings of this study are available from the corresponding author upon request.\\

%%%END OF MAIN TEXT%%%


\balance


%%%REFERENCES%%%
\bibliography{references} %You need to replace "rsc" on this line with the name of your .bib file
\bibliographystyle{rsc} %the RSC's .bst file

\end{document}